# Safety Analysis of Metasurface-Based Near-field Wireless Power Transfer System for Deep Implant

Maoyuan Li
*Dept. of Electronic Systems, Norw. Uni. of Science, and Tech Intervention Center, Oslo University Hospital,* Norway
maoyuanli.bit@gmail.com
ORCID: 0000-0002-6826-0701

Ali Khaleghi
*Dept. of Electronic Systems, Norw. Uni. of Science, and Tech Intervention Center, Oslo University Hospital,* Norway
ali.khaleghi@ntnu.no
ORCID: 0000-0002-8372-1529

Ilangko Balasingham
*Dept. of Electronic Systems, Norw. Uni. of Science, and Tech Intervention Center, Oslo University Hospital,* Norway
ilangko.balasingham@ntnu.no
ORCID: 0000-0001-5259-3221

***Abstract*— Wireless power transfer is a method for energizing future implantable medical electronics. In this study, a metasurface-based near-field magnetic wireless power transfer system for deep implants is presented, and electromagnetic safety parameters, including field distributions, specific absorption rate (SAR), and temperature variations, are evaluated. The power transfer is modeled for a receiver implant at a distance of 8.5 cm from the designed metasurface. Based on the results, a maximum localized SAR of 0.072 mW/kg is achieved when the efficiency is 1.62 %. Moreover, continuous power transfer shows that the local tissue temperature rises by less than 1.1°C.**

## I. Introduction

Currently, advancements in semiconductor technology are leading to the development of electronic device innovations that can control and treat physiological disorders. These active implantable medical devices (AIMDs) need a high power consumption density, even at a millimeter or smaller scale. Moreover, the majority of AIMDs are equipped with wireless communication modules for data transfer and device monitoring. These modules provide an additional burden on the battery life as wireless transmission requires high transmission energy due to the significant attenuation of the signals. Due to the low power density, the energy storage components will take up a sizeable percentage of the device's footprint. Energy harvesting (EH) and wireless power transfer (WPT) techniques have been demonstrated to mitigate insufficient power supplies [1].

The energy harvesting process utilizes body-generated energy, including thermal energy from body heat, chemical energy from glucose, and kinetic energy from mechanical motion [2]. However, the various transducers can only extract energy less than 10 μW. Applications requiring radios or control electronics are restricted due to the comparatively large consumption. In contrast, the WPT system for implants can transfer a provisional power of 1-10 mW. WPT techniques are adaptable to regulating the input voltage to provide steady and adequate power for the load. Given that lead- and battery-free alternatives are increasingly used in therapeutic settings [3], power transfer becomes more attractive for implants. In order to obtain milliwatt levels of power for deep-tissue (>5 cm) micro-implants, midfield powering was suggested [4]. However, the received power is limited due to tissue losses and the specific absorption rate (SAR) that limit the received power to less than 1mW for the deep implants [5].

This study suggests a metasurface-based near-field magnetic wireless power transfer system since near-field coupling features relatively low SAR. On the other hand, the metasurface (MTS) can assist the powering system in extending the operating region and compensating for misalignments. Different strategies can be adopted depending on how the MTS affects electromagnetic wave propagation. An MTS slab can be inserted beneath the *RX* or *TX* to reflect electromagnetic waves and improve coupling. The slab can also be positioned on the side to block EM waves, improving the power transfer efficiency (PTE) while reducing electromagnetic leakage or interference. To intensify the magnetic coupling, an MTS slab can be placed between the *TX* and *RX* [2]. Furthermore, magnetic distributions in tissue have no apparent discrepancies from those in free space because the quasistatic (nonradiative) magnetic fields mediate the coupling. This feature can guarantee a stable transmission link and avoid complicated antenna analysis and design.

WPT systems necessitate a comprehensive safety study for upcoming demonstrations. Using an EM-based WPT system, the body will be exposed to electromagnetic coupling [6]. In general, electromagnetic safety considers human tissue impacts, which include SAR and thermal variation. Specifically, transmission energy levels could not be higher than the SAR limit for human body tissues. To comply with safety guidelines defined by the IEEE C95.1-1999 and ICNIRP, the SAR should be confined to 1.6 W/kg for 1 g of human tissue and 2 W/kg for 10 g of human tissue [7]. Additionally, the internationally accepted safety standard of the International Electrotechnical Commission (IEC) specifies limits for the maximum tissue temperature to minimize risks [8]. It is suggested that for all people, the maximum local core temperature of any tissue should be limited to 39 °C. Overall, this study assesses SAR, the intensity of the field, and temperature rise in human tissues to determine the safety of tissue exposure.

## II. Finite Element Simulation

For optimum power transfer, the recommendation is to hold the ratio between the diameter of the *TX* and transmission distance close to unity. Thus, the dimensions of the *TX* and MTS are selected equally. The model includes a planar coil outside the body and an implant inside the biological tissues. The implant uses a helix-shaped coil within a depth inside the tissues. A 2-by-2 MTS slab is inserted between the *TX* and *RX* to alter the generated magnetic field and focus the field at the receiver coil. Thus, the PTE can be significantly improved. Figure 1 depicts the entire geometric system and the equivalent circuit.

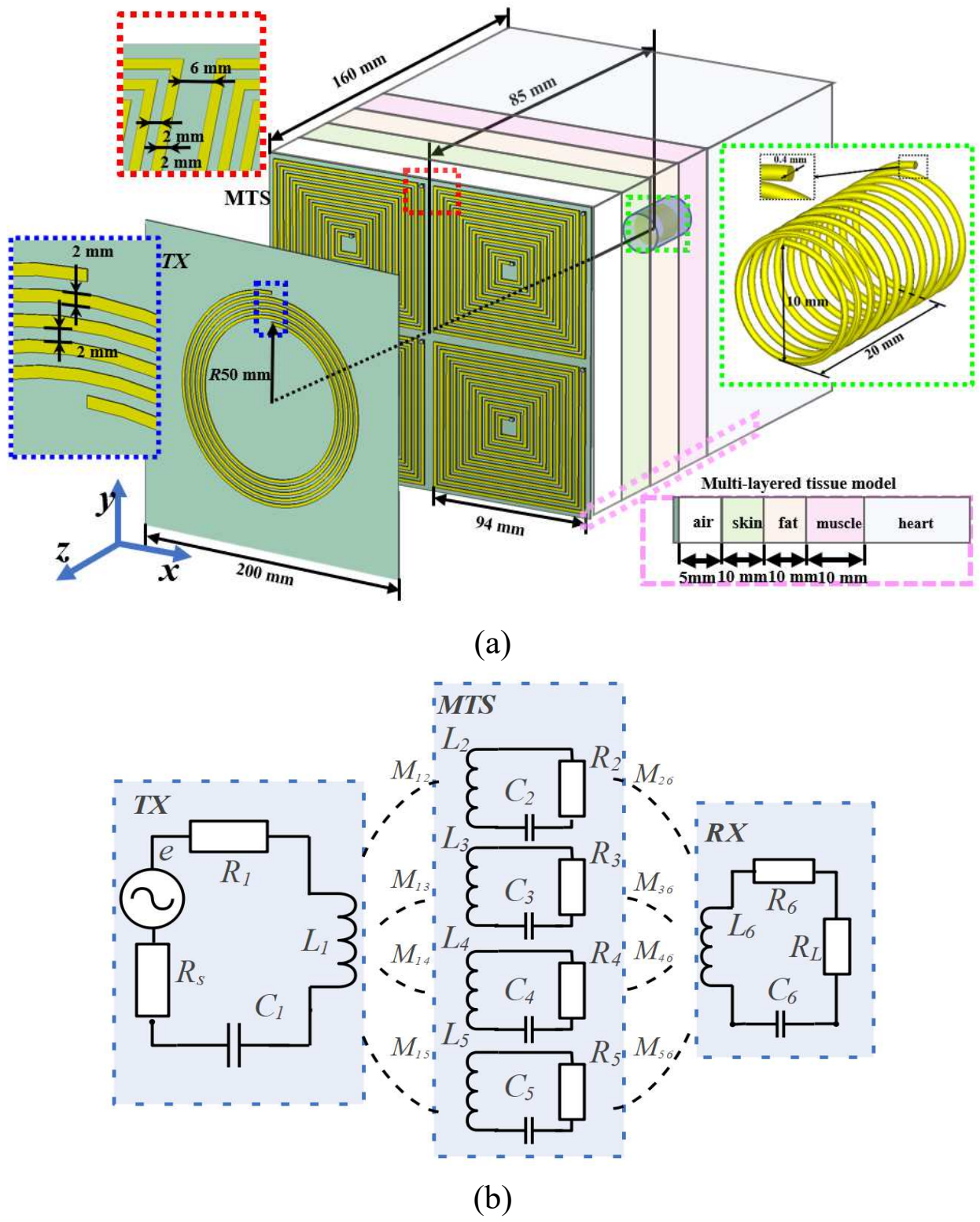

Fig. 1. (a) The geometric model of a metasurface-based near-field magnetic WPT system, and (b) the corresponding equivalent circuit.

The electromagnetic behavior of MTS relies on its resonant nature. In general, near-field MTS-based systems feature large $Q$ factors [9]. Therefore, selecting suitable external capacitances to balance the reactance of coils is essential. Moreover, we suggest a design at 6.78 MHz to follow the industrial, scientific, and medical (ISM) spectrum in the biomedical field and A4WP in the WPT standardization. The method in [10] is used to optimize the capacitance values, where we derived $C_1$=93.8pF, $C_2$=95.8 pF, and $C_6$=1.33 nF. Since the system is symmetrical, the impedance parameters in each element of MTS are the same. TABLE I provides descriptions of additional impedance parameters in Fig. 1(b).

### *A. Model Setting*

A full wave analysis is applied based on CST Microwave Studio (CST Computer Simulation Technology AG, Darmstadt, Germany). The designed WPT system consists of an active transmitter coil, a metasurface, and a passive receiver. As shown in Fig. 1(a), the transmitter is a 12 cm-diameter, identical, five-turn circular spiral coil. Lossy copper wire is used for simulation, and the thickness of the coil is 0.2 mm. MTS includes 2-by-2 rectangular planar spirals. The outside lengths of the spirals are all 94 mm. The two spirals are separated by 6 mm. The proposed element has a total inductance of 5.75 μH and a high $Q$-factor value of 163. A 10-turn helical circular inductor is created as the receiver. The length and diameter of the receiver are set to 2 cm and 1 cm, respectively (Fig. 1 (a)). Moreover, the distance between *TX* and MTS along the $z$-direction is 2 cm, and the distance between MTS and *RX* along the $z$-direction is 8.5 cm. The design parameters are listed in TABLE I.

TABLE I. GEOMETRICAL PARAMETER OF *TX*, METASUREFACE AND *RX*.

| Parameter | Transmitter | Metasurface | Receiver |
|---|---|---|---|
| Material | Lossy copper | Lossy copper | Lossy copper |
| Conductivity | 5.96×107 S/m | 5.96×107 S/m | 5.96×107 S/m |
| Inner diameter | 100 mm | 10 mm | 10 |
| Outer diameter | 144 mm | 94 mm | / |
| Thickness | 0.2 mm | 0.2 mm | / |
| Length | / | / | 20 |
| Metal radius | / | / | 0.2 |
| Turns | 5 | 10 | 10 |
| Induction | 5.28 μH ($L_1$) | 5.75 μH ($L_2$) | 0.41 μH ($L_6$) |
| Inherent resistor | 0.58 Ω ($R_1$) | 1.5 Ω ($R_2$) | 0.19 Ω ($R_6$) |
| External capacitance | 93.8pF ($C_1$) | 95.8 pF ($C_2$) | 1.33 nF ($C_6$) |
| $Q$ factor | 387 | 163 | 92 |

For simplicity and repeatability of the results, a five-layered homogenous tissue block (20×20×20 cm$^3$) is established to emulate the human environment. The Cole-Cole equation is used to fit the dielectric characteristics of the skin, fat, muscle, bone, and heart [11]. Namely,

$$\acute{\varepsilon}_r=\varepsilon_\infty+\sum_{n=1}^{4}\frac{\Delta\varepsilon_n}{1+\left(j\omega\tau_n\right)^{1-\alpha_n}}+\frac{\sigma_0}{j\omega\varepsilon_0}. \quad (1)$$

$\varepsilon_\infty$ is the permittivity at infinite frequency. The magnitude of the dispersion is described as $\Delta\varepsilon_n$. $\tau_n$ is the relaxation time, and $\sigma_0$ represents the static ionic conductivity. Therefore, the frequency-related complex permittivity, $\acute{\varepsilon}_r$, is obtained. The relative conductivity ($\sigma$ [S/m]) and permittivity ($\varepsilon_r$) are depicted in Fig. 2. *RX* is assumed to be placed in this multi-layered tissue model. The receiver is covered with an air case to prevent the induced current from diffusing into human tissue. An air gap of 10 cm around the simulation box is considered, and the open boundary is used to limit the 3D mesh of the simulation region. The peak power of 1 W is assumed to activate the transmitter in full-wave electromagnetic analysis. The *TX* and *RX* have input and load impedances of $R_s$=5 Ω and $R_L$=1 Ω, respectively. The frequency domain solver is adopted for computations. A frequency range between 4.746 and 8.814 MHz has been chosen for the simulation.

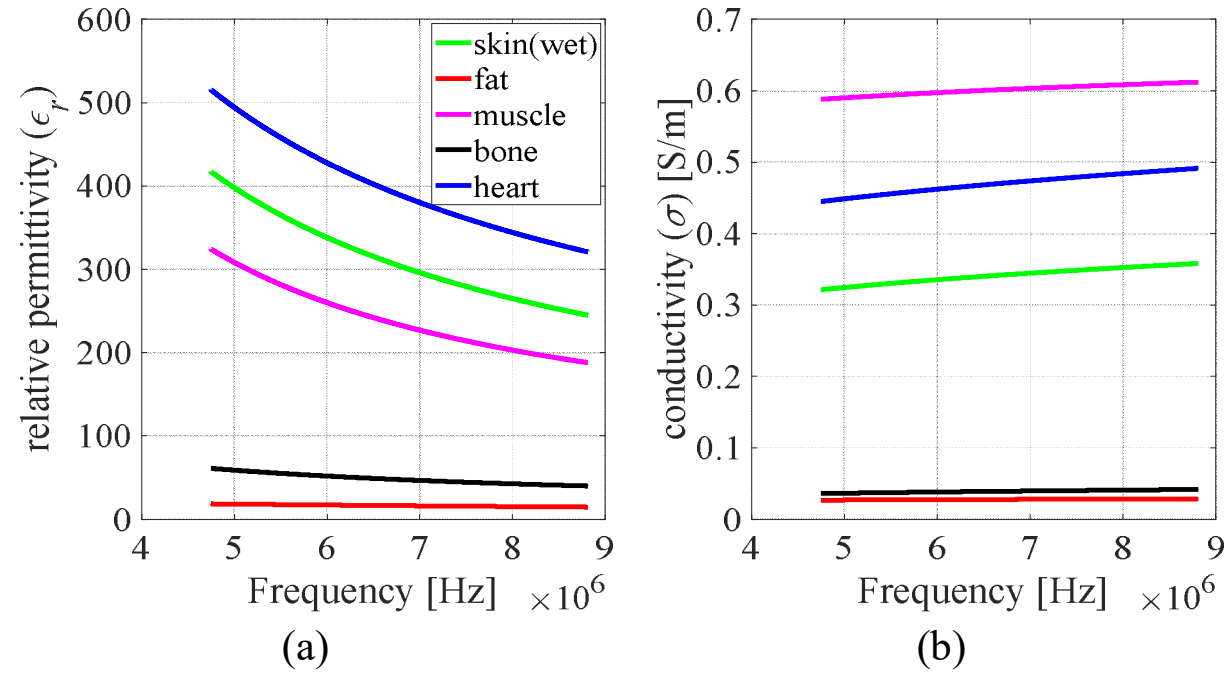

Fig. 2. (a) Relative permittivity, $\varepsilon_r$, and (b) conductivity of human tissues in the frequency range of 4.746-8.814 MHz.

## III. SIMULATION RESULTS

### *A. Coupling and Effective Permeability*

Figure 3 shows the scattering parameter over the 2-by-2 MTS near-field magnetic WPT system. According to the transmission coefficient $S_{21}$ and the reflection coefficient $S_{11}$, the coupling efficiency ($\eta$) is defined as,

$$\eta = \frac{|\bar{S}_{21}|^2}{1-|\bar{S}_{11}|^2} \tag{2}$$

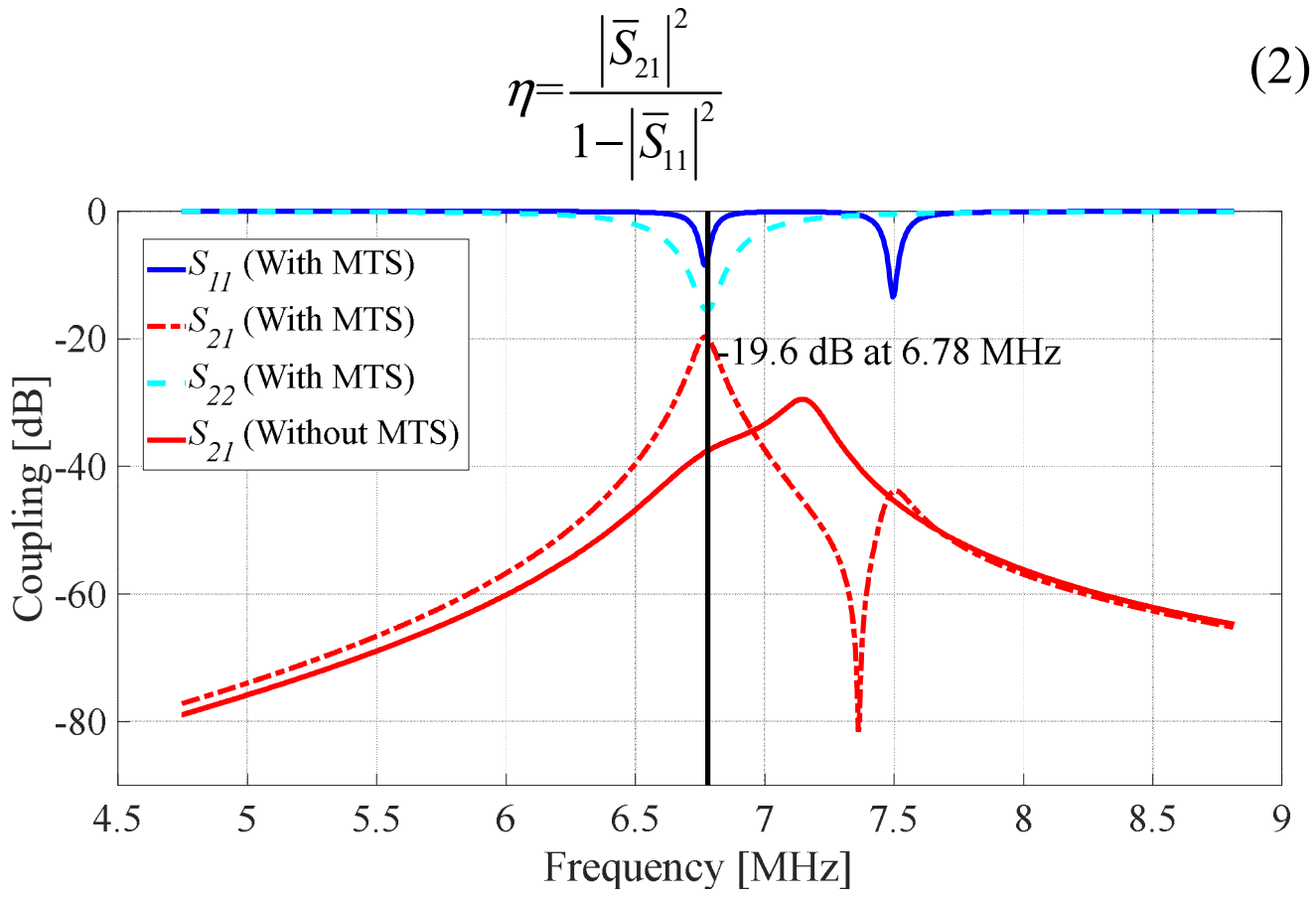

Fig. 3. The scattering parameter for the 2-by-2 MTS near-field magnetic WPT system.

As shown in Fig. 3, the derived $S_{21}$ is -19.6 dB at 6.78 MHz, resulting in an efficiency of 1.62 %. When the MTS slab is removed, the maximal transmission coefficient is -29.4 dB. It indicated that the captured power multiplies by 10 times when the MTS is used. Here, the transmitting signal excites the elements in MTS to induce the current. The same phase of excited and induced current will form to enlarge the magnetic field under resonating.

Moreover, the core concept of the metasurface is to produce materials using artificially designed and fabricated structural units to achieve the desired properties and functionalities. This artificial arrangement can tune the effective permittivity or permeability of the metamaterial to positive, near-zero, or negative values. In the system, it is imperative for the MTS to exhibit an effective negative permeability within the frequencies of interest (specifically, 6.78 MHz as per this paper) and thus to achieve magnetic field focusing. Here, the effective permeability, $\mu_r$, is derived from the aspect of the average magnetic dipole moment to avoid applying a far-field wave, where each element connects with the external capacitors, and [12],

$$\mu_r = 1 + \frac{M}{H} = 1 + \frac{\mu_0 S_{loop}^2}{L_{eff} V_{unit}} \frac{\omega^2}{-\omega^2 + j\omega(\omega_0 / Q) + \omega_0^2}, \tag{3}$$

where $M$ is averaged magnetic dipole moment per unit volume; $H$ is the external magnetic field intensity generated from the excitation; the corresponding magnetic flux density is $B=\mu_0 H$ in vacuum or air, and $L_{eff}$ is defined as effective inductance. The resonant angular frequency is represented as $\omega_0^2=1/(L_{eff}C_2)$, and the quality factor is $Q=\omega_0 L_{eff}/R$; $S_{loop}$ is the enclosed area of the element, and $V_{unit}$ is the effective volume of the element.

As a result, the effective permeability is plotted in Fig. 4. The resonance frequency is 6.65 MHz. At the wireless power transmission frequency of 6.78 MHz, the element's effective relative permeability becomes -2.72+0.56i. The negative real part enhances near-field evanescent waves. Thus, the coupling between *TX* and *RX* coils is increased, and the PTE is improved.

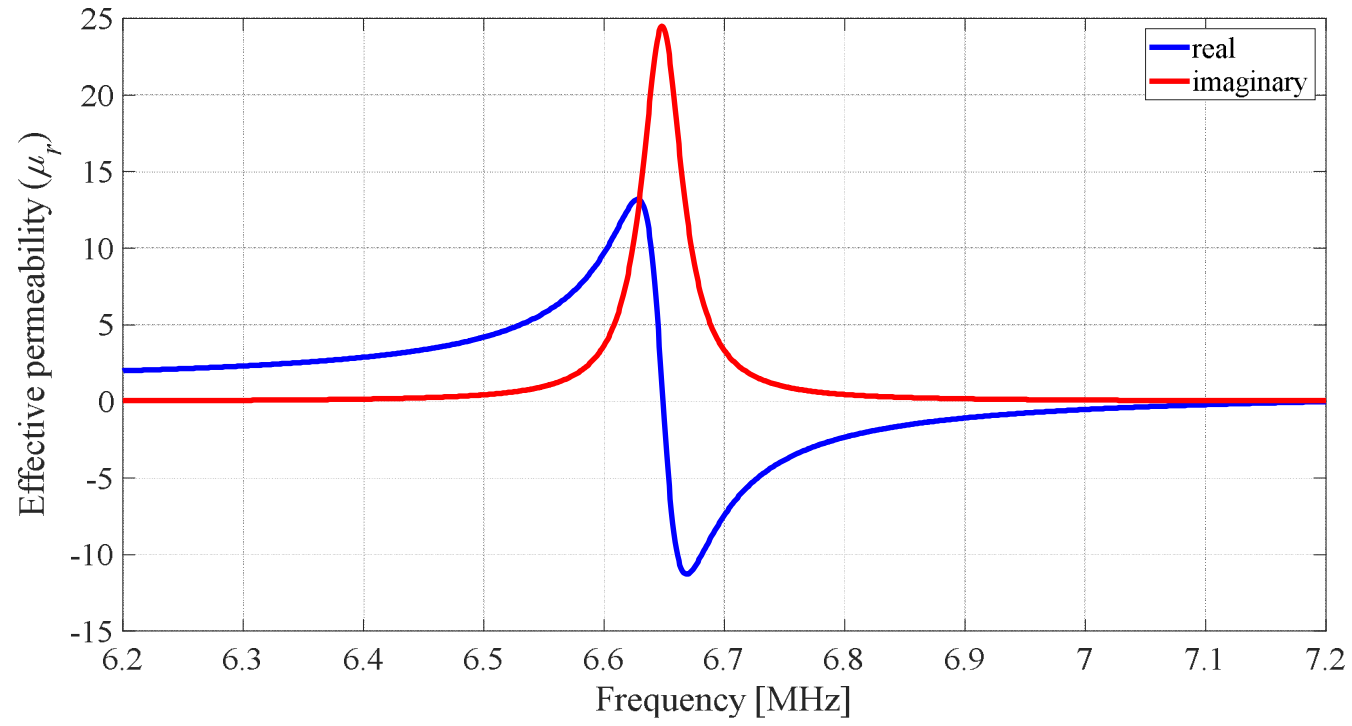

Fig. 4. Simulation results of relative permeability with metamaterial element.

## B. *The Distributions of E-, H-field, Wave Impedance*

The simulated distributions of $H$-, $E$-field, wave impedance, and SAR at 6.78 MHz are shown in Fig. 5. Compared with Fig 5 (a) and (b), in the near-field region, the electromagnetic field is decoupled. $H$-field is associated only with permeability. Because of the non-magnetic property of the human body ($\mu_r$=1), the body is transparent to the quasistatic magnetic field. The human body channel cannot fundamentally influence the magnetic field variation. Unlike permeability, the relative permittivity of human tissue, $\varepsilon_{r,\text{tissue}}$, is orders of magnitude higher than that of air ($\varepsilon_{r,\text{air}}\approx 1$). Its value drops with increasing frequency. As wavelength $\lambda^2 \propto 1/\varepsilon_r$, higher $\varepsilon_{r,\text{tissue}}$ implies a lower $\lambda$ compared to air, i.e. $\lambda_{Body} << \lambda_{air}$. Such a high discontinuity in $\lambda$ can give rise to the confinement of $E$-field. As shown in Fig. 5 (b), human tissue blocks the majority of $E$-field density outside. Moreover, the simulated wave impedance is less than 6 Ω, as shown in Fig. 5 (c). It indicates that the $H$-field dominates the wireless power transfer [13]. The small $E$-field imposes a small SAR.

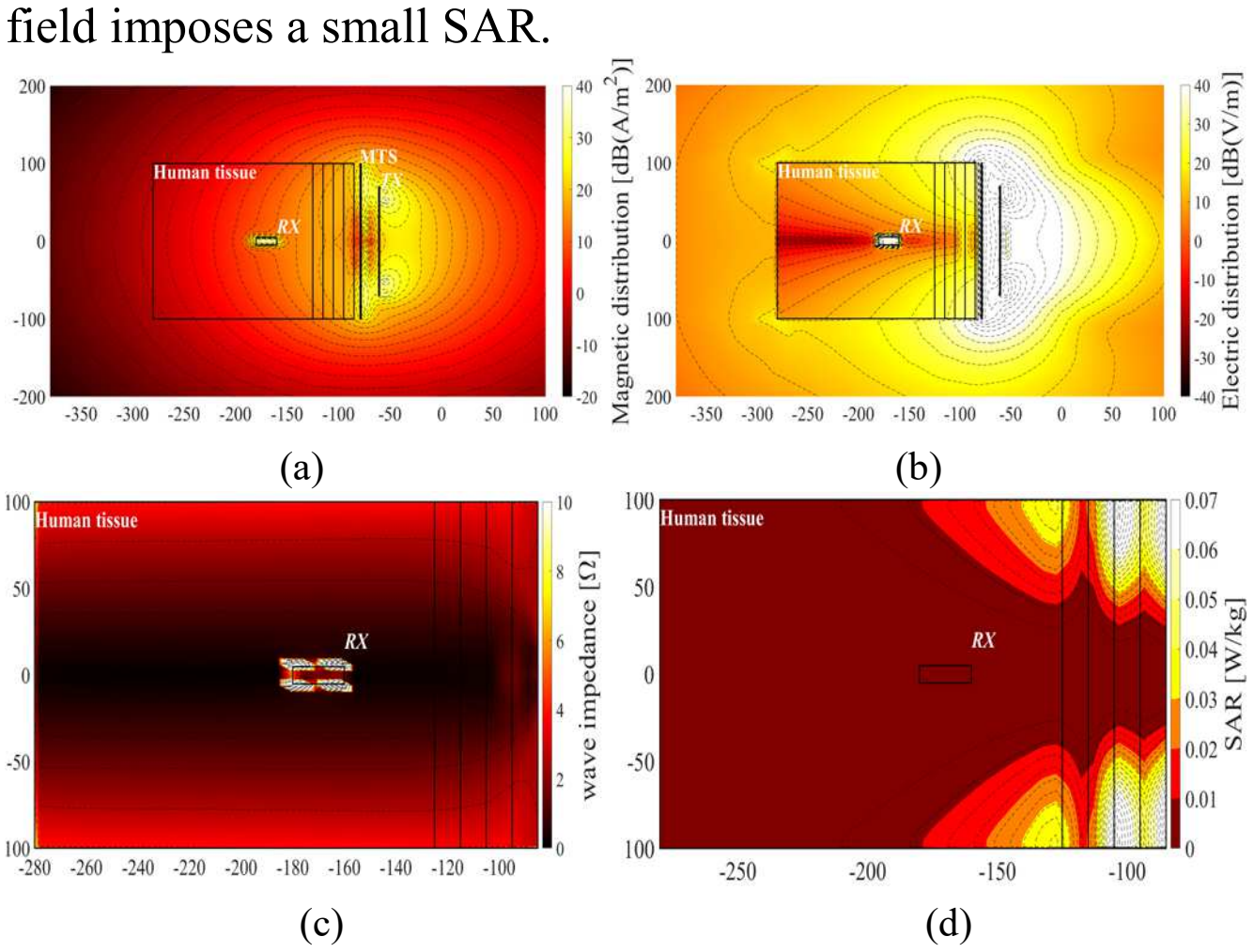

Fig. 5. The distribution of (a) $H$-, (b) $E$-field, (c) wave impedance, and (d) SAR.

## C. *Safety Assessment Regarding SAR*

The near-field metasurface-based WPT system produces a more localized electromagnetic field and SAR in the body. This condition prompts particular emphasis on evaluating the partial-body exposure rather than the whole-body exposure considered in larger resonant-type systems. The basic restrictions for occupational exposure specified by ICNIRP are evaluated in TABLE II. Here, SAR is determined by the dissipated power in a tissue volume divided by its mass,

$$SAR=\frac{1}{\Delta V}\int_{\Delta V}\frac{\sigma(\vec{r})}{\rho(\vec{r})}\left|E(\vec{r})\right|^{2}dr, \tag{4}$$

where $\sigma$ is the sample electrical conductivity; $E$ is the RMS electric field; $\rho$ is the sample density; $\Delta V$ is the volume of the sample. When the frequency of 6.78 MHz is operated, the maximum localized SAR is less than 0.072 $W\cdot kg^{-1}$, which is far below the limit value, as shown in Fig. 5 (d). Moreover, ICNIRP requires the reference peak value of $2.7\times10^{-4}f$ $V\cdot m^{-1}$ from 100 kHz to 10 MHz for localized exposure to electromagnetic fields. The maximal electric density at 6.78 MHz is only 29.8 V/m, which meets the specified requirements [7].

TABLE II. ICNIRP RESTRICATIONS AND THE SIMLATED SAR

| | ICNIRP | this work |
|---|---|---|
| Frequency rang | 100 kHz-10 MHz | 4.746-8.814 MHz |
| Whole-body average SAR | 0.4 $W\cdot kg^{-1}$ | < 0.072 $W\cdot kg^{-1}$ |
| Localized SAR (head and trunk) | 10 $W\cdot kg^{-1}$ | |
| Localized SAR (limbs) | 20 $W\cdot kg^{-1}$ | |
| Induced electric field intensity | $2.7\times10^{-4}f$ (1830.6 V/m, $f$=6.78 MHz) | < 29.8 V/m |

### D. Temperature Rise

In this study, electrical-thermal co-simulation using CST is conducted. The steady thermal solver is used. The situation represents a static temperature distribution of the human body under charging at an infinite time. Here, the effect of blood flow and the metabolic heating rate are considered in the simulation process. The ambient temperature is set at 300 K (26.85 ℃). The blood temperature is set as 310.15 K (37 ℃). The temperature change because of the induced SAR can be obtained by solving the Pennes' bioheat equation [14].

$$\rho c\frac{\partial T}{\partial t}=\nabla\cdot\left(k\nabla T\right)+\rho Q_{met}+\rho\left(SAR\right)-B\left(T-T_{blood}\right), \tag{5}$$

$$B=\rho_{blood}c_{blood}\rho\omega,$$

where the $\rho$ [kg/m$^3$] is the material density, $c$ is the specific heat capacity, $k$ is the thermal conductivity, $Q_{met}$ is the metabolic heat generation rate, $B$ is the blood perfusion coefficient, $\omega$ is the blood perfusion rate, and $T_{blood}$ is blood temperature. The corresponding parameter has already been integrated into CST. The thermal parameters of human tissues are listed in TABLE III. As shown in Fig. 6, the maximal temperature in the human body is 37.73 °C during the excited power of *TX* is set at 5 W in this thermal simulation, which will not cause harm to the human body.

TABLE III. THERMAL PARAMETERS OF HUMAN TISSUES.

| Tissue properties (unit) | Skin | Fat | Muscle | Bone | Heart |
|---|---|---|---|---|---|
| Density (kg/m$^3$) | 1100 | 900 | 1080 | 2000 | 1080 |
| Specific heat capacity (W/m/K)) | 3500 | 2500 | 3500 | 1300 | 3700 |
| Thermal conductivity (W/K/m) | 0.3 | 0.2 | 0.5 | 0.4 | 0.5 |
| Metabolic rate (W/m$^3$) | 2000 | 300 | 500 | 600 | 10000 |

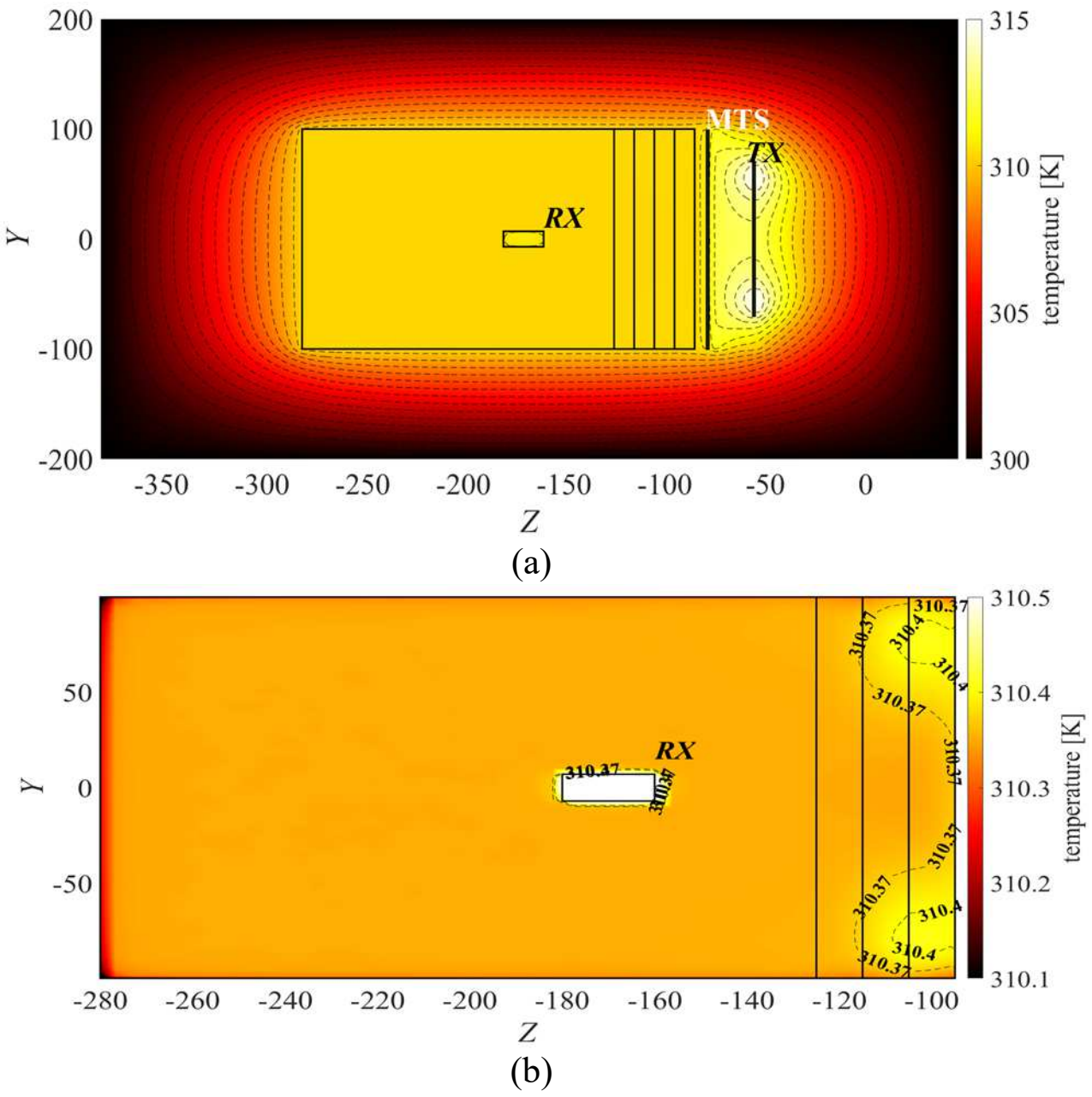

Fig. 6. Thermal rise of (a) the whole estabilised environment and (b) the human tissue, where the larger thermal variation is located the surface of skin closely.

Moreover, a relationship between input power and temperature/SAR has been simulated in thermal simulation. SAR and temperature rise are positively associated with input power improvement. In the simulated example, even after increasing the input power to 10 W, the temperature change (red line) and SAR (green line) remained within the safety limit, yielding received power of 162 mW. According to the simulation findings, the suggested method demonstrates the superiority of received power over the existing WPT systems, and received power is at least an order of magnitude higher than the state-of-the-art works. However, the governing equation has not considered the convective heat transfer coefficient. The core temperature of *TX* (closed loop) gets relatively high as the input power increases, as shown in Fig. 7.

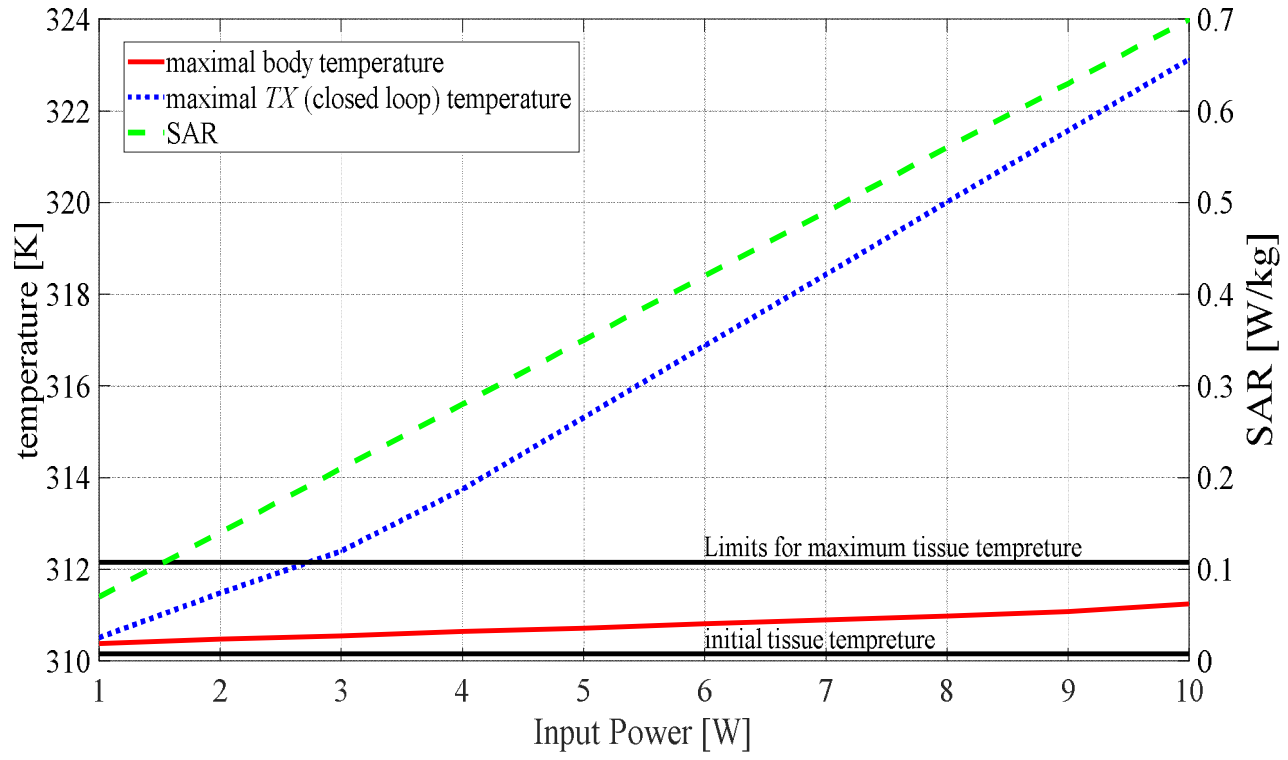

Fig. 7. The relationship between input power and temperture/SAR.

## IV. CONCLUSION

In this paper, a metasurface-based near-field magnetic wireless power transmission system for deep implants is proposed. The corresponding finite element method simulation is conducted. This wireless power transfer system has undergone a safety analysis in preparation for future demonstrations. SAR and temperature variation are considered in the safety assessment. The power transfer is modeled for a receiver implant separated from the metasurface by 8.5 cm. The efficiency can achieve 1.62 %. The results show that the maximal simulated values of specific absorption rate (SAR, 0.072 W/kg @ 1W) and temperature variation (1.1°C @ 10 W) remain in the safety range. The system suggests a promising solution for deep-tissue implant powering issues.

## ACKNOWLEDGMENT

This work was funded in part by the Research Council of Norway project, Communication Theoretical Foundation of Wireless Nanonetworks (grant no. 287112), by the South-Eastern Norway Regional Health Authority project, Pre-Clinical Studies for Multi-node Leadless Cardiac Pacemaker (grant no. 2020049), and by Horizon 2020 FET-OPEN project, B-CRATOS, (grant no. 965044).